\documentclass[aps,prb,showpacs,intlimits,amsmath,amssymb,twocolumn]{revtex4-1}
\usepackage{graphicx}
\usepackage{bm}

\begin{document}

\title{Nonequilibrium dynamical mean-field theory for the charge-density-wave phase of the Falicov-Kimball model}
\author{O.~P.~Matveev$^{1,2}$, A.~M.~Shvaika$^2$, T.P.~Devereaux$^{3,4}$, and 
J.~K.~Freericks$^1$}
\affiliation{$^1$ Department of Physics, Georgetown University, Washington, DC
20057, USA}
\affiliation{$^2$ Institute for Condensed Matter Physics of the National Academy of Sciences of Ukraine,
Lviv, 79011 Ukraine}
\affiliation{$^3$ Geballe Laboratory for Advanced Materials, Stanford University, 
Stanford, CA 94305,USA}
\affiliation{$^4$ Stanford Institute for Materials and Energy Sciences (SIMES), 
SLAC National Accelerator Laboratory, Menlo Park, CA 94025, USA }

\begin{abstract}
Nonequilibrium dynamical mean-field theory (DMFT) is developed for the case of the charge-density-wave ordered phase. We consider the spinless Falicov-Kimball model which 
can be solved exactly. This strongly correlated system is then placed in an uniform external dc electric field. We present a complete derivation for nonequilibrium 
dynamical mean-field theory Green's functions defined on the Keldysh-Schwinger time contour. We also discuss numerical issues involved in solving the coupled equations.
\end{abstract}

\pacs{71.10.Fd, 71.45.Lr, 72.20.Ht}
\maketitle

\section{Introduction}
There are a number of strongly correlated materials that have charge-density-wave (CDW) behavior. Static order occurs in the transition metal di- and trichalchogenides, 
which display either quasi one dimensional (NbSe$_3$) or quasi two dimensional (TaSe$_2$ or TbTe$_3$) order~\cite{NbSe3,TaSe2}. Three-dimensional charge-density-wave 
order is observed in Ba$_{1-x}$K$_{x}$BiO$_3$ compounds~\cite{cdw_exp}. There is a longstanding question concerning the nature of the CDW order, namely, whether the 
order is driven electronically, with a lattice instability following the electronic instability, or {\it vice versa}. Indeed, in real materials the charge and lattice 
degrees of freedom are usually strongly coupled, but recent time-resolved core-level photoemission spectroscopy experiments~\cite{hellmann} for some CDW materials 
indicate an electronically driven nature to ordering. This makes it reasonable to study the CDW phase for strongly correlated electronic systems that do not include a 
coupling to the lattice. 

While most theoretical interests in strongly correlated systems have been concentrated on equilibrium behavior, recent experiments on pump-probe spectro\-scopy~\cite{hellmann,petersen} 
have caused an increase of attention to non\-equilibrium dynamics. These experiments display a non\-equilibrium melting of the CDW state, which is manifested by a 
filling of the gap in the photoemission spectrum, while the order parameter remains nonzero. This phenomenon has been examined with an exactly solvable model~\cite{shen_fr1} 
for a system starting at zero temperature.

We use the Falicov-Kimball model in our analysis because it is one of the simplest models~\cite{falicov_kimball} which possesses static charge-density-wave ordering 
and has an exact solution within DMFT~\cite{brandt_mielsch1} (for a review see Ref.~\cite{freericks_review}). The many-body formalism 
for nonequilibrium dynamical mean-field theory is straightforward to develop within the Kadanoff-Baym-Keldysh formalism~\cite{kadanoff,keldysh}. Since the many-body 
perturbation theory diagrams are topologically identical for both equilibrium and non\-equilibrium perturbation theories~\cite{langreth}, the perturbative analysis of 
Metzner~\cite{metzner} guarantees that the nonequilibrium self-energy remains local. The basic structure of the iterative approach to solving the DMFT
equations~\cite{jarrell} continues to hold. Detailed development of the nonequilibrium DMFT approach has been done for the case of the uniform phase of the Falicov-Kimball 
model~\cite{frturzlat}. Here we generalize this method to the case of the CDW ordered phase.

\section{Static order and the Hamiltonian}
\label{sec:1}
In order to describe the CDW ordered state, one has to rewrite the Hamiltonian assuming the existence of the charge modulation. This can be done in two ways: by introducing 
two sublattices ``$A$'' and ``$B$'' in real space or by the nesting of the Brillouin zone (BZ) at the modulation vector $\mathbf{Q} =(\pi, \pi, \dots)$ in a reciprocal space. 
The modulation vector $\bf Q$ defines the sublattices by
\begin{align}
e^{i\mathbf{Q}\mathbf{R}_i}=\begin{cases}
                                 1, & \quad \mathbf{R}_i\in A, \\
                                -1, & \quad \mathbf{R}_i\in B.
                            \end{cases} 
\end{align}
In the ordered phase, due to nesting of the Fermi surface, the BZ is reduced and instead of the annihilation (creation) operators with momentum $\bf k$ defined in the initial 
BZ by $c_{\bf{k}}=\dfrac{1}{N}\sum_i e^{i\bf{k}\bf{R}_i}c_i$, one has to introduce two fermionic operators in momentum space in the reduced zone ($\bf k \in \textrm{rBZ}$) as 
$\tilde{c}_{1\mathbf{k}} = c_{\mathbf k}$ and $\tilde{c}_{2\mathbf{k}} = c_{\mathbf{k+Q}}$.

Now, one can write down the relations between annihilation (creation) operators defined on the sublattices ($A,B$) and in the rBZ ($1,2$)
\begin{align}
\tilde{c}_{1\mathbf{k}} &= \dfrac{1}{N}\sum_{i\in A}e^{i\mathbf{k}\mathbf{R}_i}c_i+\dfrac{1}{N}\sum_{i\in B} e^{i\mathbf{k}\mathbf{R}_i}c_i = \frac{c_{\mathbf{k}A}+c_{\mathbf{k}B}}{\sqrt{2}},\\
\tilde{c}_{2\mathbf{k}} &= \dfrac{1}{N}\sum_{i\in A}e^{i(\mathbf k+\mathbf Q)\mathbf{R}_i}c_i+\dfrac{1}{N}\sum_{i\in B} e^{i(\mathbf k+\mathbf Q)\mathbf{R}_i}c_i 
= \frac{c_{\mathbf{k}A}-c_{\mathbf{k}B}}{\sqrt{2}}, 
\nonumber
\end{align}
or in a matrix form as follows: 
\begin{align}
\label{transformation}
   \begin{bmatrix}
  \tilde{c}_{1\mathbf k} \\
  \tilde{c}_{2\mathbf k}
   \end{bmatrix}
=\hat{U} \begin{bmatrix}
  c_{\mathbf k A} \\
  c_{\mathbf k B}
   \end{bmatrix}, \quad\text{where}\quad
\hat{U}= \begin{Vmatrix}
  \dfrac{1}{\sqrt{2}} &  \dfrac{1}{\sqrt{2}} \\
  \dfrac{1}{\sqrt{2}} & -\dfrac{1}{\sqrt{2}}
   \end{Vmatrix}.
\end{align}

We use both sublattice ($A,B$) and rBZ ($1,2$) bases. Any two-operator-pro\-duct-type quantity, \textit{e.g.} the one-particle Green's function, can be defined with the additional 
sublattice indices $\mathcal{\hat{O}}(\mathbf{k})=\|\mathcal{O}_{\alpha,\beta}(\mathbf{k})\|$ $(\alpha,\beta=A,B)$, or with the rBZ indices
$\mathcal{\hat{\widetilde{O}}}(\mathbf{k})=\|\mathcal{O}_{m,n}(\mathbf{k})\|$ $(m,n=1,2)$, and these representations are connected by the aforementioned unitary transformation
\begin{align}
\mathcal{\hat{\widetilde{O}}}(\mathbf{k})=\hat{U} \mathcal{\hat{O}}(\mathbf{k}) \hat{U}^{-1}.
\end{align}
We work in units where $\hbar=c=e=a=1$.

The time-dependent Hamiltonian of the spinless Fa\-li\-cov-Kimball model on a bipartite lattice has the form
\begin{align}
\mathcal{H}(t)&=\sum_{i\alpha}(U n^\alpha_{id} n^\alpha_{if}-\mu_d n^\alpha_{id})
-\sum_{ij\alpha\beta} t^{\alpha\beta}_{ij}(t)c^{\dag}_{i\alpha} c_{j\beta}.
\end{align}
We consider the case when charged fermions interact with an external electric field which is spatially uniform. 
This allows us to describe the electric field via a time-dependent vector potential in the Coulomb gauge as $\mathbf{E}(t)=-d\mathbf{A}(t)/dt$. Interaction with this external 
field results in a Peierls' substitution to the kinetic term of the Hamiltonian.

In the sublattice representation ($A,B$), the local part of the Hamiltonian is diagonal and the non-local kinetic one is off-diagonal in the case of the nearest-neighbor 
hopping. In the rBZ representation ($1,2$) it is {\it vice versa}.

The Fourier transformation to momentum space gives the time-dependent kinetic term in the form 
\begin{align}
\hat{\mathcal{H}}_{kin}(t) &= \sum_{k}\begin{bmatrix}
c^{\dag}_{\mathbf{k}A} & c^{\dag}_{\mathbf{k}B}
\end{bmatrix}\hat{\epsilon}(\mathbf k-\mathbf{A}(t))\begin{bmatrix}
c_{\mathbf{k}A}\\ c_{\mathbf{k}B}
\end{bmatrix}
\nonumber \\
&= \sum_{k}\begin{bmatrix}
\tilde{c}^{\dag}_{1\mathbf{k}} & \tilde{c}^{\dag}_{2\mathbf{k}}
\end{bmatrix}\hat{\tilde{\epsilon}}(\mathbf k-\mathbf{A}(t))\begin{bmatrix}
\tilde{c}_{1\mathbf{k}}\\ \tilde{c}_{2\mathbf{k}}
\end{bmatrix},
\end{align}
where an extended band energy $\hat{\epsilon}(\mathbf k-\mathbf{A}(t))$\cite{frturzlat} in matrix form in the ($A,B$) basis becomes
\begin{align}\label{eps}
&\hat{\epsilon}(\mathbf k-\mathbf{A}(t))= \\
&\begin{Vmatrix}
 \scriptstyle 0 & \scriptstyle\epsilon(\mathbf k)\cos (A(t))+\bar{\epsilon}(\mathbf k)\sin (A(t)) \\[1ex]
 \scriptstyle \epsilon(\mathbf k)\cos (A(t))+\bar{\epsilon}(\mathbf k)\sin (A(t)) & \scriptstyle 0
 \end{Vmatrix}.
 \nonumber
\end{align}
In the rBZ representation ($1,2$), an extended band energy matrix is diagonal. Here, we introduced the band energies 
$\epsilon(\mathbf k)=\lim\limits_{d\rightarrow\infty}\dfrac{-t^*}{\sqrt{d}}\sum\limits_{r=1}^d \cos k_r$ and 
$\bar{\epsilon}(\mathbf k)=\lim\limits_{d\rightarrow\infty}\dfrac{-t^*}{\sqrt{d}}\sum\limits_{r=1}^d \sin k_r$,
and we set $t^*$ as our energy unit. 

\section{Real time Green's function in the CDW phase}
\label{sec:2}
The key object of our interest is the time-dependent Green's function that is defined on the Keldysh-Schwin\-ger contour\cite{kadanoff,keldysh,frturzlat} 
\begin{align}
G^{c}_{\bf k}(t,t')=-i\langle\mathcal{T}_{c} c_{\bf k}(t)c_{\bf k}^{\dag}(t')\rangle.
\end{align}
We start from a formal solution of Dyson's equation for the lattice Green's function:
\begin{align}
\label{glat}
\hat{G}^c_{\epsilon,\bar{\epsilon}}(t,t')=\biggl{[}(\hat{G}^{c,non}_{\epsilon,\bar{\epsilon}})^{-1}-\hat{\Sigma}^c\biggr{]}^{-1}(t,t').
\end{align}
In the two-sublattice representation ($A,B$), the self-energy $\hat{\Sigma}^c(t,t')$ is diagonal and the noninteracting Gre\-en's function $\hat{G}^{c,non}_{\epsilon,\bar{\epsilon}}(t,t')$ 
is non-diagonal [because of extended band energy $\hat{\epsilon}(\mathbf k-\mathbf{A}(t))$ in Eq.~(\ref{eps})]. But in the rBZ representation ($1,2$) 
$\hat{G}^{c,non}_{\epsilon,\bar{\epsilon}}(t,t')$ becomes diagonal  and its analytical expression is known from the uniform solution\cite{frturzlat}. We apply the unitary 
transformation in Eq.~(\ref{transformation}) to the self-energy $\hat{\Sigma}^c(t,t')$ and we find the solution for the lattice Green's function in Eq.~(\ref{glat}) 
in explicit matrix form in the rBZ basis ($1,2$) as follows: 
\begin{align}
\label{mglat}
&\hat{\tilde{G}}^c_{\epsilon,\bar{\epsilon}}(t,t')= \\
&\begin{Vmatrix}
 \scriptstyle[(G^{c,non}_{\epsilon,\bar{\epsilon}})^{-1}-\frac{\Sigma^{c,A}+\Sigma^{c,B}}{2}](t,t') & \scriptstyle-\frac{\Sigma^{c,A}-\Sigma^{c,B}}{2}(t,t') \\
 \scriptstyle-\frac{\Sigma^{c,A}-\Sigma^{c,B}}{2}(t,t') & \scriptstyle[(G^{c,non}_{-\epsilon,-\bar{\epsilon}})^{-1}-\frac{\Sigma^{c,A}+\Sigma^{c,B}}{2}](t,t')
\end{Vmatrix}^{-1}\nonumber
\end{align}

The Green's functions and self-energies defined on the Keldysh-Schwinger contour are continuous matrix operators of two time variables. We discretize the contour with 
several different grids and then find the continuous matrix limit using Lagrange's interpolation formula. To find the inverse matrix in Eq.~(\ref{mglat}), where its components 
are matrices in time variables, we need to apply the block matrix pseudo-inverse formula. 

We have performed a transition onto the BZ basis in order to find the lattice Green's function in terms of the noninteracting Green's function. Further, we  construct the 
other DMFT equations to make the system of equations self-consistent. In our case of a CDW phase, it is more convenient and clear how to do this in the sublattice basis ($A,B$). 
Hence, at this point we apply an inverse transformation from Eq.~(\ref{transformation}) onto the sublattice basis and write down expressions for the components of the lattice 
Green's function in the  $(A,B)$ basis in terms of its components in the rBZ basis.

The next step in the DMFT approach is to find the local Green's function, which is further mapped onto a single impurity problem. Since we are considering a two-sublattice 
system, we have two local Green's functions for the $A$ and $B$ sublattices, respectively. 

To calculate the local Green's functions, we need to sum over the reduced BZ all $(\mathbf k-\mathbf{A}(t))$-dependent functions. Because of presence of the time-dependent vector 
potential, one has to replace the summation over the BZ by a double integration over the energies
\begin{align}
 \hat{G}^c_{loc}(t,t')=\dfrac{1}{N}\sum\limits_{\mathbf k}\hat{G}^c_{\mathbf{k}}(t,t')
 =\int d\epsilon\int d\bar{\epsilon} \rho(\epsilon,\bar{\epsilon})\hat{G}^c_{\epsilon,\bar{\epsilon}}(t,t')
\end{align}
with a joint density of states which is a Gaussian in each variable for the hypercubic lattice \cite{frturzlat} and is given by 
$\rho(\epsilon,\bar{\epsilon})=\text{exp}(-\epsilon^{2}-\bar{\epsilon}^{2})/\pi$.

\begin{figure}
 \includegraphics[width=0.48\textwidth]{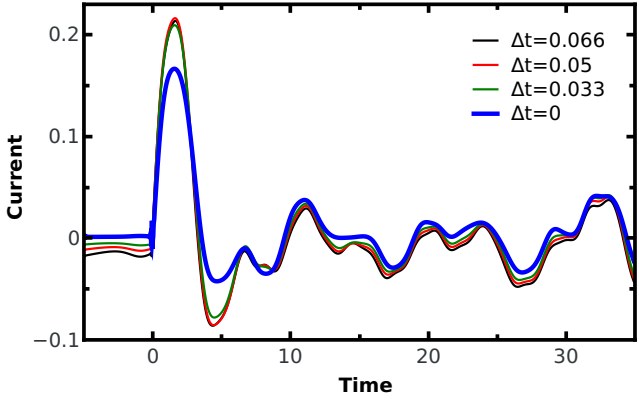}
 \caption{Quadratic extrapolation of the current for $U=1.0$ with $E=1$ at temperature $T=0.033$ ($\Delta n_f=0.495$). Different curves correspond to different $\Delta t$ values.}
 \label{current}
\end{figure}

Finally, we find the components of the local Green's function for each sublattice as follows:

\begin{align} \label{gloc}
&G^{c,A}_{loc}(t,t')
 =\int \int d\epsilon d\bar{\epsilon} \rho(\epsilon,\bar{\epsilon})[I+\Lambda\Sigma][I-K\Sigma\Lambda\Sigma]^{-1}K, \nonumber \\
&G^{c,B}_{loc}(t,t')
 =\int \int d\epsilon d\bar{\epsilon} \rho(\epsilon,\bar{\epsilon})[I-\Lambda\Sigma][I-K\Sigma\Lambda\Sigma]^{-1}K, 
\end{align}
where we introduced new quantities $K$, $\Lambda$, and $\Sigma$ which satisfy
\begin{align}\label{kls}
K&= 
 \biggl\{(G^{c,non}_{\epsilon,\bar{\epsilon}})^{-1}-\dfrac{\Sigma^{c,A}+\Sigma^{c,B}}{2}\biggr\}^{-1}(t,t'), \nonumber \\
\Lambda&= 
 \biggl\{(G^{c,non}_{-\epsilon,-\bar{\epsilon}})^{-1}-\dfrac{\Sigma^{c,A}+\Sigma^{c,B}}{2}\biggr\}^{-1}(t,t'), \nonumber \\
\Sigma&=
 \dfrac{\Sigma^{c,A}-\Sigma^{c,B}}{2}(t,t').
\end{align}

Next, we need to map the local lattice Green's function onto the impurity Green's function. Employing Dyson's equation, we introduce an effective medium   
\begin{equation}\label{g0}
 \hat{G}^{c}_{loc}(t,t')=[(\hat{G}^{c}_0)^{-1}-\hat{\Sigma}^{c}]^{-1}(t,t')=\hat{G}^{c}_{imp}(t,t'). 
\end{equation}
The effective medium $\hat{G}^{c}_0(t,t')$ is diagonal in the sublattice representation and its components are equal to 
\begin{align}
 G^{c,\alpha}_0(t,t')=[(G^{c,\alpha}_{loc})^{-1}+\Sigma^{c,\alpha}]^{-1}(t,t'), (\alpha=A,B)
\end{align}

\begin{figure}
 \includegraphics[width=0.48\textwidth]{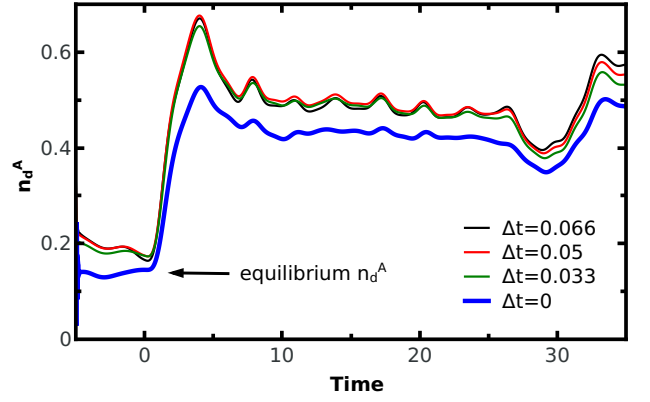}
 \caption{Quadratic extrapolation of $n_d^A$ for $U=1.0$ with $E=1$ at temperature $T=0.033$ ($\Delta n_f=0.495$). Different curves 
correspond to different $\Delta t$.}
 \label{conc}
\end{figure}

On the other hand, it can be found from the Dyson's equation that defines an effective dynamical mean field $\hat{\lambda}^{c}(t,t')$.
Its components for each sublattice equal to:
\begin{align}
G^{c,\alpha}_{0}(t,t')=[(i{\partial_t}^c+\mu_d)\delta_c(t,t')-\lambda^{c,\alpha}(t,t')]^{-1}.
\end{align}
We next extract the dynamical mean fields for each sublattice 
\begin{align}\label{lambda}
&\lambda^{c,\alpha}(t,t')=(i{\partial_t}^c+\mu_d)\delta_c(t,t')-(G^{c,\alpha}_{0})^{-1}(t,t') \\
&=(i{\partial_t}^c+\mu_d)\delta_c(t,t')-(G^{c,\alpha}_{loc})^{-1}(t,t')-\Sigma^{c,\alpha}(t,t'),
\nonumber
\end{align}
which are the effective fields for the nonequilibrium single-impurity problems.

Now, we close the system of DMFT equations with the solution of the impurity problem, which is known for the Falicov-Kimball model: 
\begin{align}\label{gimp}
 &G^{c,\alpha}_{imp}(t,t')=[(1-n_f^\alpha)G^{c,\alpha}_0+n_f^\alpha G^{c,\alpha}_1](t,t'), \quad \text{where} \nonumber \\
 &G^{c,\alpha}_1(t,t')=[(1-G^{c,\alpha}_0 U)]^{-1}G^{c,\alpha}_0](t,t').
\end{align}

The difference between the $A$ and $B$ sublattices is defined by the order parameter $\Delta n_f$, which is equal to the difference of the $f$-particle occupations at different 
sublattices ($2\Delta n_f=n_f^A-n_f^B$). In the CDW phase, the total concentration of localized electrons is fixed, and $\Delta n_f$ is defined from the initial equilibrium 
condition. In nonequilibrium, the order parameter remains the same as in equilibrium because the $f$-particles of the Falicov-Kimball model do not interact with the external 
field.

We calculate the current in the Hamiltonian gauge by evaluating the operator average
\begin{align}
 \langle\mathbf{j}(t)\rangle=-i\sum\limits_{\mathbf{k}}\mathbf{v}(\mathbf{k}+\theta(t)\mathbf{E}t)G_{\mathbf{k}}^<(t,t), 
\end{align}
where the lesser Green's function $G_{\mathbf{k}}^<(t,t)$ is extracted from contour ordered Green's function, and the velocity component is 
$v_i(\mathbf{k})=\lim\limits_{d\rightarrow\infty} t^{*}\sin(\mathbf{k}_i)/\sqrt{d}$.  

The results of the numerical calculation depend stro\-ng\-ly on the discretization of the time interval and we need to take the limit of $\Delta t\rightarrow 0$. Smaller $\Delta t$ 
results in larger matrices so we exploit an extrapolation procedure to get more accurate results. In Fig. \ref{current}, we present 
the results for the current at $U=1.0$ that corresponds to a metallic phase at temperature $T=0.033$ and order parameter $\Delta n_f=0.495$. Different curves correspond to 
different values of $\Delta t$ and we use a quadratic Lagrange's interpolation formula to extrapolate the result to $\Delta t=0$. These show the correct zero current before 
the field is turned on at $t=0$ (we choose $E=1$) and their longer time behavior agrees with zero temperature calculations for the Falicov-Kimball model as well\cite{shen_fr1}. Similarly, in 
Fig. \ref{conc}, we show the results for concentration of $d$-electrons on the $A$-sublattice. Here, extrapolation shows an even more significant shift of the nonzero $\Delta t$ 
results and its result is in a good agreement with the $T=0$ calculations\cite{shen_fr1}. Other results for more values of the parameters will be presented elsewhere.

\acknowledgments

This work was supported by the Department of Energy, Office of Basic Energy Sciences, Division of Materials Sciences and Engineering under Contract Nos. DE-AC02-76SF00515 
(Stanford/SIMES), DE-FG02-08ER46542 (Georgetown) and DE-SC0007091 (for the collaboration). Computational resources were provided by the National Energy Research Scientific 
Computing Center supported by the Department of Energy, Office of Science, under Contract No. DE- AC02-05CH11231. J.K.F. was also supported by the McDevitt bequest at Georgetown.

\end{document}